\documentclass[aps,prl,twocolumn,showpacs,preprintnumbers,amsmath,amssymb,superscriptaddress]{revtex4-1}
\usepackage{graphicx}% Include figure files
\usepackage{dcolumn}% Align table columns on decimal point
\usepackage{bm}% bold math
\newcommand{\nc}{\newcommand}
\nc{\be}{\begin{equation}}
\nc{\ee}{\end{equation}}
\nc{\bea}{\begin{eqnarray}}
\nc{\eea}{\end{eqnarray}}
\nc{\bean}{\begin{eqnarray*}}
\nc{\eean}{\end{eqnarray*}}
\nc{\mb}{\mbox}
\nc{\rnc}{\renewcommand}
\nc{\vk}{\mb{\bf k}}
\nc{\vp}{\mb{\bf p}}
\nc{\vn}{\mb{\bf n}}
\nc{\vq}{\mb{\bf q}}
\nc{\rr}{\mb{\bf r}}
\nc{\vz}{\hat {\mb{\bf z}}}
\nc{\vj}{\mb{\boldmath$j$}}
\nc{\vg}{\mb{\boldmath$g$}}
\nc{\x}{\mb{\boldmath$x$}}
\nc{\A}{\mb{\boldmath$A$}}
\nc{\va}{\mb{\boldmath$a$}}
\nc{\vs}{\mb{\boldmath$\sigma$}}
\nc{\vpi}{\mb{\boldmath$\pi$}}
\nc{\nab}{\nabla}
\nc{\X}{\sf x}

\begin{document}

\title{Spin Josephson effects in Exchange coupled Anti-ferromagnets}

\author{Yizhou Liu}
\affiliation{Department of Electrical and Computer Engineering, University of California, Riverside, CA 92521, USA}
\affiliation{Center of Spins and Heat in Nanoscale Electronic Systems, University of California, Riverside, CA 92521, USA}

\author{Gen Yin}
\affiliation{Department of Electrical and Computer Engineering, University of California, Riverside, CA 92521, USA}
\affiliation{Center of Spins and Heat in Nanoscale Electronic Systems, University of California, Riverside, CA 92521, USA}

\author{Jiadong Zang}
\affiliation{Department of Physics and Material Science Program, University of New Hampshire, Durham, New Hampshire 03824, USA}

\author{Roger Lake}
\thanks{Email: rlake@ece.ucr.edu}
\affiliation{Department of Electrical and Computer Engineering, University of California, Riverside, CA 92521, USA}
\affiliation{Center of Spins and Heat in Nanoscale Electronic Systems, University of California, Riverside, CA 92521, USA}

\author{Yafis Barlas}
\thanks{Email: yafisb@ucr.edu}
\affiliation{Center of Spins and Heat in Nanoscale Electronic Systems, University of California, Riverside, CA 92521, USA}
\affiliation{Department of Physics and Astronomy, University of California, Riverside, CA 92521, USA}

%--Outline
% 1) Introduction.
% 2) Coupled magnetization dynamics in AFMIs.
% 3) Persistent a.c spin current due to spin injection.  
% 4) Microscopic calculation of the inter-layer exchange.
% 5) Conclusion and outlook.

\begin{abstract}
The energy of exchange coupled antiferromagnetic insulators (AFMIs) 
is a periodic function of the relative in-plane orientation of the 
N\'{e}el vector fields. 
We show that this leads to oscillations in the relative magnetization of 
exchange coupled AFMIs separated by a thin metallic barrier. 
These oscillations pump a spin current ($I_{S}$) through the metallic spacer 
that is proportional to the rate of change of the relative in-plane 
orientation of the N\'{e}el vector fields. 
By considering spin-transfer torque induced by a spin chemical potential ($V_{S}$) 
at one of the interfaces, 
we predict non-Ohmic $I_{S}$-$V_{S}$ characteristics of AFMI exchange 
coupled hetero-structures, 
which leads to a non-local voltage across a spin-orbit coupled metallic spacer. 
\end{abstract}


\maketitle

{\em Introduction}---The discovery of the giant magneto-resistance~\cite{GMR1,GMR2} along with spin transfer torques~\cite{JS_stt_1996, berger_stt_1996} has led to emergence of spintronics as a platform for modern electronic devices.
Most conventional spintronics applications rely on transport of spin-polarized electrons, which is typically accompanied by Joule heating.   
Additionally, fast spin relaxation of the conduction electron in metals and semi-conductors makes these applications challenging~\cite{RevModDS}.  
A promising alternative is to combine elements of conventional spintronics to generate and manipulate pure spin-currents carried by the collective excitations of magnetically ordered insulators~\cite{SZhangmagnondrag}.
The effects of magnon mediated spin currents have been detected experimentally in hetero-structures with ferromagnetic insulators~\cite{kajiwara_transmission_2010, Shi_Pt-YIG-Pt}.

A recent proprosal for a more exotic method of transporting spin 
harnesses the ground states of both easy-plane FMs~\cite{AllanSSa,yaroslav_FM,AllanSSb} 
and antiferromagnetic insulators (AFMIs)~\cite{takei_superfluid_2014}.
It has been long appreciated that magnetically ordered systems with spontaneously broken 
$U(1)$ symmetry generically support metastable spin spiral states that can 
transfer spin angular momentum without dissipation~\cite{HalperinHohenberg_69,Sonin}.
These spin super-currents are analogues of charge or mass supercurrents in 
superconductors and superfluids~\footnote{This analogy is useful even in the absence of strict conservation laws for spin, as long as the violation of the conservation laws is weak}.
However, only recent advancements in material growth have made it possible to 
detect dissipationless spin transport in appropriately designed heterostructures.
In this regard, heterostructures composed of AFMIs are advantageous to those composed of FMs, 
as they are less sensitive to stray fields or dipolar interactions, 
which can destroy dissipationless spin transport in easy-plane FMs~\cite{Brataas_SF}. 

\begin{figure}
\begin{center}
\includegraphics[width=3.5in,height=1.5in]{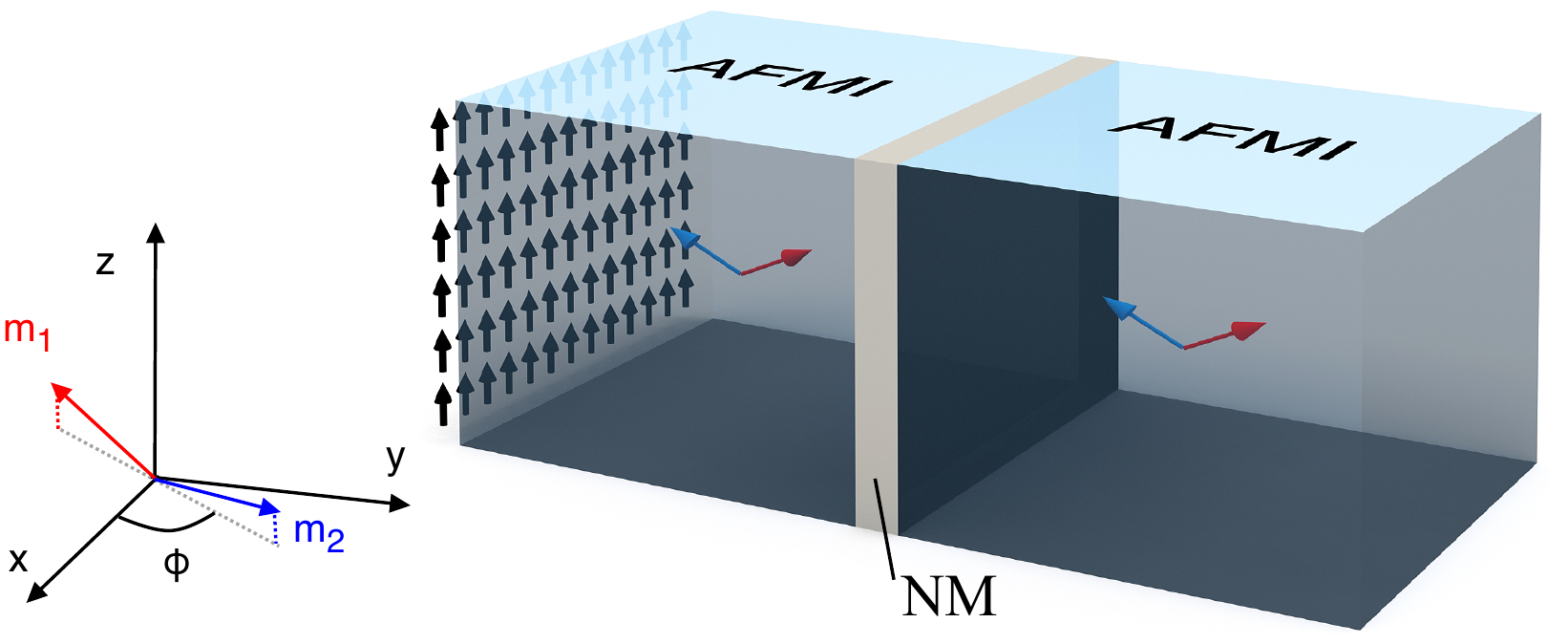}
\caption{Schematic diagram of the proposed heterostructure to detect the Josephson effect 
in spin superfluids. 
The heterostructure consists of two antiferromagnetic insulators (AFMIs) separated by a 
thin metallic spacer. 
The magnetization of the AFMIs lies in the $xy$-plane as indicated in the inset 
showing the direction of the N\'{e}el vector and phase $\phi$, 
with a spin canting in the $\hat{z}$-direction. 
A spin chemical potential of up spins on the left interface of the AFMI can drive 
an oscillating spin current through the metallic spacer via spin pumping. 
The spin current flowing through a spin-orbit (SO) coupled metallic spacer can be 
detected via the inverse spin Hall effect.}
\label{fig:structure}
\end{center}
\end{figure}

Coupled superfluids and superconductors exhibit oscillations in the mass 
or charge current on the application of a constant chemical potential, called the
a.c. Josephson effect~\cite{josephson_1962}.
This effect occurs because the energy of coupled superfluids and superconductors is a periodic function 
of their relative phase difference.
Here, we show that the spin superfluid analogy can be further extended to realize 
Josephson-like oscillations of the spin currents flowing through exchange coupled 
antiferromagnetic insulators (AFMIs).
Since magnetic insulators only support spin currents and couple to spin chemical potentials, 
a framework based on the hydrodynamic theory of spin waves and spin transfer 
torques is developed.
This oscillatory spin current can be detected by injecting a pure spin current on the left side 
of the heterostructure illustrated in Fig.~\ref{fig:structure}, 
which consists of two AFMIs separated by a thin metallic spacer.
As we explain, a spin chemical potential established perpendicular to the direction of 
the N\'{e}el vector field, drives an oscillatory spin supercurrent which can be converted 
to a charge current via the inverse spin Hall effect through a metallic spacer with 
large spin-orbit coupling.
The key observation that motivates this proposal is that the energy of 
exchange coupled AFMIs is a periodic function of the 
relative in-plane orientation of the N\'{e}el vector fields. 

{\em Coupled magnetization dynamics}---
Consider the heterostructure in
Fig.~\ref{fig:structure}, consisting of
two bipartite lattice AFMIs separated by a thin nonmagnetic metallic spacer that
provides a local interlayer exchange coupling between the two AFMIs. 
Each AFMI has a staggered spin orientation 
$\textbf{s}_{i} (\textbf{r}) = S_{i} (\textbf{r})/S$ 
where $i = \pm$ denotes the left (right) AFMI
and $S$ is the saturated spin density.
The long-wavelength effective Hamiltonian describing the fluctuation of the 
AFMIs can be expressed in terms of two continuum fields 
$\textbf{n}_{i}(\textbf{r})$ (the N\'{e}el vector field) and 
$\textbf{m}_{i}(\textbf{r})$ (the canting field), 
with the local spin orientation 
$\textbf{s}_{i} (\textbf{r})= \eta_{i,\textbf{r}} \textbf{n}_{i} (\textbf{r}) 
\sqrt{1-|\textbf{m}_{i}(\textbf{r})|^2} + \textbf{m}_{i} (\textbf{r})$
with the constraints 
$| \textbf{n}_{i} | = 1$ and  $\textbf{n}_{i} \cdot \textbf{m}_{i} = 0$, 
where $\eta_{i,\textbf{r}} = \pm 1$ for the $A(B)$ sublattices 
\cite{haldane_nonlinear_1983}.
Assuming that the N\'{e}el vectors lie in the $xy$-plane with an interlayer exchange interaction 
$\sum_{\textbf{r},\textbf{r}'} J_{\textbf{r},\textbf{r}'} \textbf{s}_{L,\textbf{r}} \cdot \textbf{s}_{R,\textbf{r}'}$ \cite{bruno_theory_1995}, 
the effective Hamiltonian capturing the long-wavelength dynamics of this system is,
\begin{eqnarray}
\label{eq_Hamiltonian}
\mathcal{H} &=& \frac{1}{\mathcal{V}}\int d\textbf{r} \sum_{i= \pm 1} \bigg[ \frac{\rho}{2} \big(\nabla \textbf{n}_{i}(\textbf{r}) \big)^2 + \frac{\lambda}{2}  \textbf{m}^2_{i}(\textbf{r}) \\ \nonumber 
&+&\frac{J}{2} \textbf{n}_{i}(\textbf{r})\cdot \textbf{n}_{-i}(\textbf{r}) + 
\frac{J}{2} \textbf{m}_{i}(\textbf{r})\cdot \textbf{m}_{-i}(\textbf{r}) \bigg], 
\end{eqnarray}
where $\mathcal{V}$ is the volume, $J$ is the inter-layer exchange coupling of the two AFMIs, $\lambda > 0$ is the homogenous AFMI exchange coupling, and $\rho$ is the spin stiffness assumed equal for both AFMIs \cite{auerbach_interacting_1994}.
The energy of each AFMI is independent of the direction of the N\'{e}el vector $\textbf{n}_{i}$ indicating $U(1)$ symmetry, and $\lambda >0 $ ensures that $\textbf{m}_{i} =0 $ in equilibrium.
%Eq.~\ref{eq_Hamiltonian} is valid for small variations about the N\'{e}el ordered state $ |\textbf{m}_{i}| \ll |\textbf{n}_{i}|$. 
%

The long wavelength dynamics of the isolated system can be captured by the 
Landau-Lifshitz-Gilbert(LLG) equations, 
which subjected to the AFMI constraints, 
can be expressed as, 
\begin{eqnarray}
\label{LLG_n}
\hbar \dot{\textbf{n}}_{i} &=& \lambda \textbf{m}_{i} \times \textbf{n}_{i} + J  \textbf{m}_{-i} \times \textbf{n}_{i} - \hbar \alpha \textbf{n}_{i} \times \dot{\textbf{m}}_{i},  \\
\label{LLG_m}
\hbar \dot{\textbf{m}}_{i} &=& \rho \textbf{n}_{i} \times \nabla^{2} \textbf{n}_{i} + J \textbf{n}_{i} \times \textbf{n}_{-i}  - \hbar\alpha \textbf{n}_{i} \times \dot{\textbf{n}}_{i}, 
\end{eqnarray}
where ($\dot{\textbf{m}}_{i},\dot{\textbf{n}}_{i})$ denote the time derivatives 
of the fields $(\textbf{m}_{i}, \textbf{n}_{i})$, $\alpha$ is the damping constant 
assumed the same for both AFMIs, 
and henceforth we neglect the spatial dependence of the fields ($\nabla^2 \textbf{n}_{i} \sim 0$).
To implement the AFM constraints in the above equation we define: $\textbf{n}_{i} = (\cos\theta_{i} \cos \phi_{i}, \cos\theta_{i} \sin\phi_{i}, \sin\theta_{i})$, where $\phi$ is the azimuthal angle, $\theta$ is the relative angle to the xy-plane and $\textbf{m}_{i} = (-m_{\theta,i} \sin \theta_{i} \cos \phi_{i} - m_{\phi,i} \sin \phi_{i}, -m_{\theta,i} \sin \theta_{i} \sin \phi_{i} + m_{\phi,i} \cos \phi_{i}, m_{\theta,i} \cos \theta_{i})$.
With these substitutions the long-wavelength dynamics of the coupled AFMIs can be described in terms of a pair of canonically conjugate fields $(m_{\theta,i}, \phi_{i})$ and $(m_{\phi,i}, \theta_{i})$ for both AFMIs.
For small variations about the N\'{e}el ordered state $\theta_{i} \approx 0$, the LLG equations for $(m_{\theta,i},\phi_{i})$ neglecting the quadratic terms, are decoupled from the $(m_{\phi,i}, \theta_{i})$ fields, reducing to, 
\begin{equation}
\begin{aligned}
\hbar \dot{m}_{\theta, i} =  J \sin(\phi_{i} - \phi_{-i}) - \hbar \alpha \dot{\phi_{i}}, \\
\hbar \dot{\phi}_{i} = \lambda m_{\theta, i} + J m_{\theta,-i} + \hbar \alpha \dot{m}_{\theta,i}. \\
\label{Eq_magnetization_dynamics}
\end{aligned}
\end{equation}

For zero damping ($\alpha = 0$), the above equations describing the magnetization 
dynamics for exchange coupled systems AFMIs are remarkable similar to the 
Josephson equations of coupled superconductors.
This becomes evident after defining, 
$m_{\theta} = m_{\theta,L} - m_{\theta,R} $ and 
$\phi = \phi_{L} - \phi_{R}$.  
Then, Eqs. (\ref{Eq_magnetization_dynamics}) give 
$\hbar \dot{m}_{\theta} = 2 J \sin(\phi)$ and 
$\hbar \dot{\phi} = (\lambda-J) m_{\theta}$~\footnote{We restrict our analysis to small variations in the fields, for larger fluctuations a full treatment for the LLG equations must be performed. This will be discussed elsewhere.}.
The time dynamics of the relative phase is governed by $ \ddot{\phi} =  \omega_{0}^2 \sin(\phi)$, 
where the characteristic frequency $\omega_{0} = \sqrt{2 J(\lambda-J)}/\hbar$, depends on the nature of the inter-layer exchange of the coupled AFMIs.
The equation describing the phase dynamics resembles the equation of a 
simple pendulum with tilt angle $\phi$ or equivalently the motion of a 
particle with unit mass moving in a potential $U(\phi) = \omega_{0}^2 \cos(\phi)$.
This mechanical analogue provides
an intuitive understanding of the rich magnetization dynamics of 
Eqs. (\ref{Eq_magnetization_dynamics}).
\begin{figure}[t]
\begin{center}
\includegraphics[width=3.4in]{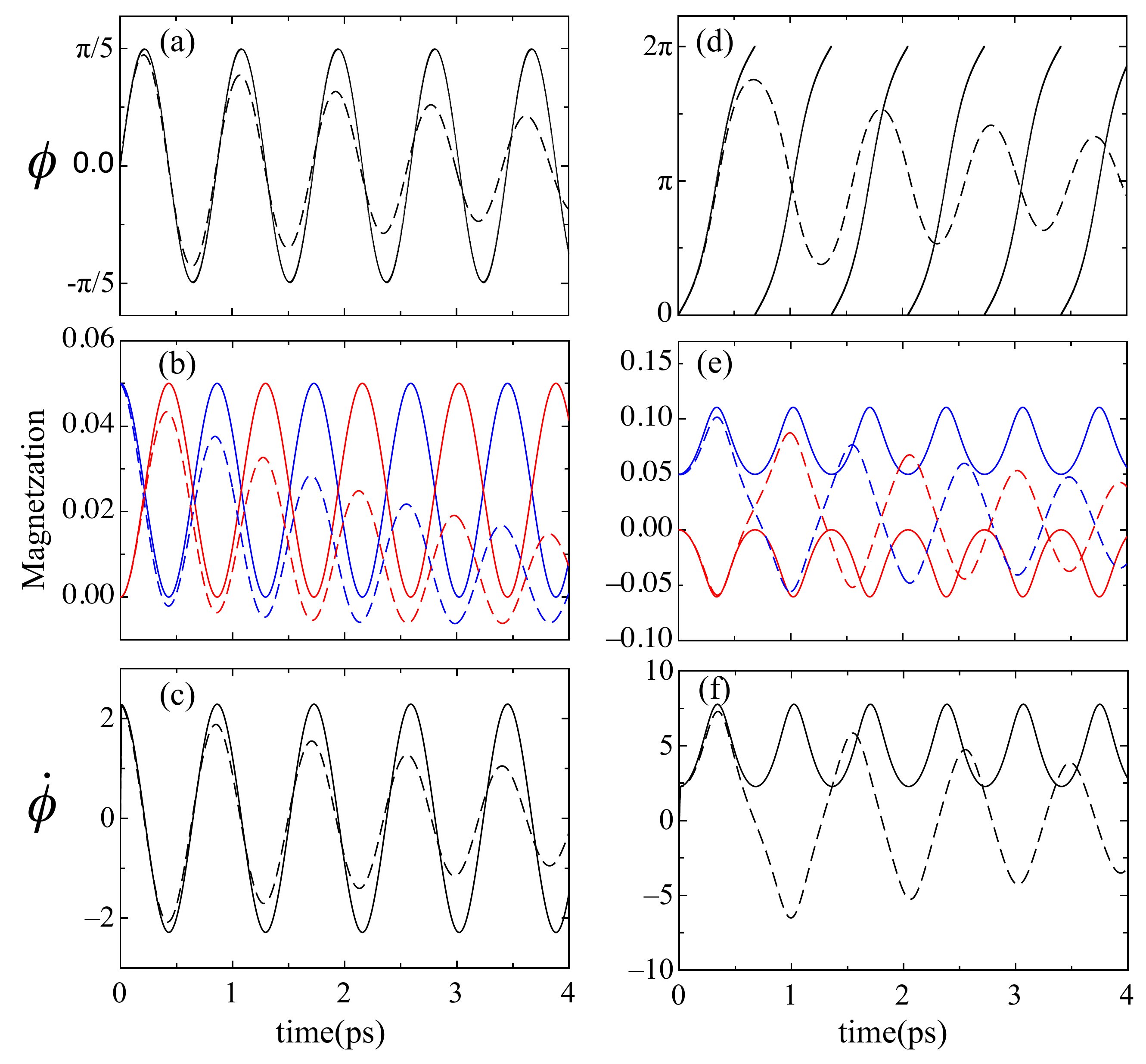} 
\caption{
Dynamics of $m_{L}$ (blue), $m_{R}$ (red), 
$\phi$, and the total spin current $I_{S}/I_{S,0}= \dot{\phi}$ 
with the initial conditions $m_{0} = 0.05$ and $\phi =0$. 
(a)-(c) are for the case of a FM inter-layer exchange $J < 0$, 
and (d)-(f) represent the case for an AFMI inter-layer exchange $J > 0$. 
All dashed lines represent the dynamics for non-zero damping with $\alpha=0.05$. 
The equations are solved for $|J|/\lambda = 1/300$ and $\lambda = 30$ meV. }
\label{fig:solution}
\end{center} 
\end{figure} 

Starting with initial conditions $\phi(t=0)=0$ and 
$m_{\theta,0} = 0.05 \propto \dot{\phi} (t =0)$, 
we solve Eqs. (\ref{Eq_magnetization_dynamics}) for both FM and AFM 
interlayer exchange coupling $J$.
The magnetization exhibits periodic oscillations with frequencies $\omega \sim 1 - 10$ THz, 
as indicted in Fig.~\ref{fig:solution}, 
with different dynamics for the FM and AFM inter-layer exchange.
Fig.~\ref{fig:solution}(a) shows the periodic variation of the phase dynamics 
for an FM exchange $|J|/\lambda= 1/300 $ with $\lambda = 30 meV$, 
which oscillates about the equilibrium point $\phi =0$. 
The magnitude of the oscillations depends on the initial velocity 
$\dot{\phi}_{0} \propto (\lambda - J) m_{\theta,0}/\hbar$.
The individual magnetizations ($m_{\theta, L}$ and $m_{\theta,R}$) 
also exhibits coupled periodic oscillations and as indicated in Fig~\ref{fig:solution} (b), 
and the total magnetization is conserved for the isolated system in the absence of any damping.  
When the initial magnetization is above a critical value $m > m_{0,c} = 2\sqrt{2|J|/(\lambda - J)}$, 
the N\'{e}el vector performs full rotations in the $xy$-plane, 
this critical value corresponds to an initial angular velocity of the pendulum 
$\dot{\phi}_{c} \geq 2|\omega_{0}|$.
The effect of the Gilbert damping with $\alpha = 0.05$
denoted by the dotted lines in Fig.~\ref{fig:solution}, results in an 
exponential damping of the phase in time, which in turns leads to an 
exponential damping of the total magnetization.

The mechanical analogue for the antiferromagnetic interlayer exchange coupling 
$J > 0$ corresponds to a simple pendulum with the initial condition 
$\phi(t=0) = \pi$, or the $\pi$-phase Josephson junction in superconductors.
For this case, as shown in Fig.~\ref{fig:solution}(d), irrespective of the initial conditions, 
the N\'{e}el vector performs complete rotations in the $xy$-plane. 
$m_{\theta,L}$ and $m_{\theta,R}$ oscillate in-phase leading to an oscillation 
in $m_{\theta}$ which is conserved in the absence of damping.  
With damping, ($\alpha = 0.05$), the individual magnetizations decay exponentially 
as indicated by the dashed lines in Fig. \ref{fig:solution}, 
and as before the total magnetization decays exponentially. 
The oscillations in the relative magnetization generate a spin current 
through the metallic spacer, as we show next.  

{\em Oscillatory spin-current}---
The oscillations in the relative magnetization induced by the dynamics in the N\'{e}el vector fields of the AFMIs pump a spin current through the metal spacer.
This spin current can be expressed as $I_{S}= I_{S,L} -I_{S,R} $, 
where  $I_{S,i} = \hbar \frac{G_{r}}{4\pi} (\textbf{n}_{i} \times \dot{\textbf{n}}_{i} + \textbf{m}_{i} \times \dot{\textbf{m}}_{i}) - \hbar \frac{G_{im}}{4\pi} (\dot{\textbf{m}}_{i})$, 
with $I_{S,L} (I_{S,R})$ defined as the spin current injected from left (right) side 
of the metallic spacer, 
$G = \mathcal{A} g^{\uparrow \downarrow} /N S$ is the spin-mixing conductance 
at the AFMI/spacer interface, $\mathcal{A}$ is the interface area and $g^{\uparrow \downarrow} = g^{\uparrow \downarrow}_{r} + i g^{\uparrow \downarrow}_{im} $ is the spin-mixing conductance per unit area \cite{cheng_spin_2014, tserkovnyak_enhanced_2002} and $N=\mathcal{V}/a_{0}^{3}$ denotes the total number of spins.
Restricting to the $\hat{z}$-component of the spin, the spin current pumped into the metallic spacer by the AFMIs can be expressed as,
\begin{equation} 
\label{spincurrent}
\begin{aligned}
I_{S} = \hbar \frac{G_{r}}{4\pi} \dot{\phi_{L}} - \hbar \frac{G_{r}}{4\pi} \dot{\phi_{R}} = \hbar \frac{G_{r}}{4\pi} \dot{\phi},
\end{aligned}
\end{equation}
valid for $|m_{i}| \ll |n_{i}|$. 
for simplicity, we assume the spin-mixing conductance is real and equal 
at both AFMI$|$spacer interfaces.   

The normalized $I_{S}/I_{S,0}$ spin current
flowing through metallic spacer, 
where $I_{S,0} = \hbar G \omega_{0}/(4 \pi )$ is the characteristic spin current 
supported by the junction, 
is plotted in Figs.~\ref{fig:solution}(c) and (f) for the FM and the AFM inter-layer exchange.
Eq.~\ref{spincurrent} states that spin current is proportional to the 
rate of change of the relative phase, 
different form the Josephson voltage phase relation in superconductor.
Similarly, we anticipate that the spin chemical potential must act, 
via spin transfer torque, 
as a source term for the rate of change of the relative magnetization.
%
%This indicates a reversal in the role of spin currents ($I_{S} \to V$) 
%and spin chemical potentials ($V_{S} \to I$), 
%when compared to Josephson effect in superconductors.
%

{\em Steady state spin-current}---Non-equilibrium spin accumulation at the left 
interface of the first AFMI, 
via the spin hall effect\cite{sinova_spin_2015} or anomalous Hall effect, 
can transfer angular momentum by inducing a spin transfer torque on the coupled AFMI system.
The spin-transfer torque can be expressed as 
$\tau_{S} = G_{r}/(4\pi) \textbf{n} \times \bm{\mu_{S}} \times \textbf{n} 
+  G_{im}/(4\pi) \bm{\mu_{S}} \times \textbf{n}$ 
where 
$\bm{\mu_{S}} = \bm{\mu_{0}} - \hbar \textbf{n} \times \dot{\textbf{n}}$ 
denotes the total non-equilibrium spin accumulation at the left interface, 
$\bm{\mu_{0}}$ is the spin accumulation, and 
$\hbar \textbf{n} \times \dot{\textbf{n}}$ 
denotes the spin pumping back-action due to the precession of the N\'{e}el vector, 
satisfying Onsager reciprocity \cite{takei_superfluid_2014}.
This non-equilibrium spin accumulation leads to a non-zero relative magnetization 
via spin-transfer torque, 
resulting in the precession of the N\'{e}el vector field 
that drives an oscillatory spin current through the metallic spacer, which we analyze next.
\begin{figure}[t]
\begin{center}
\includegraphics[width=3.4in]{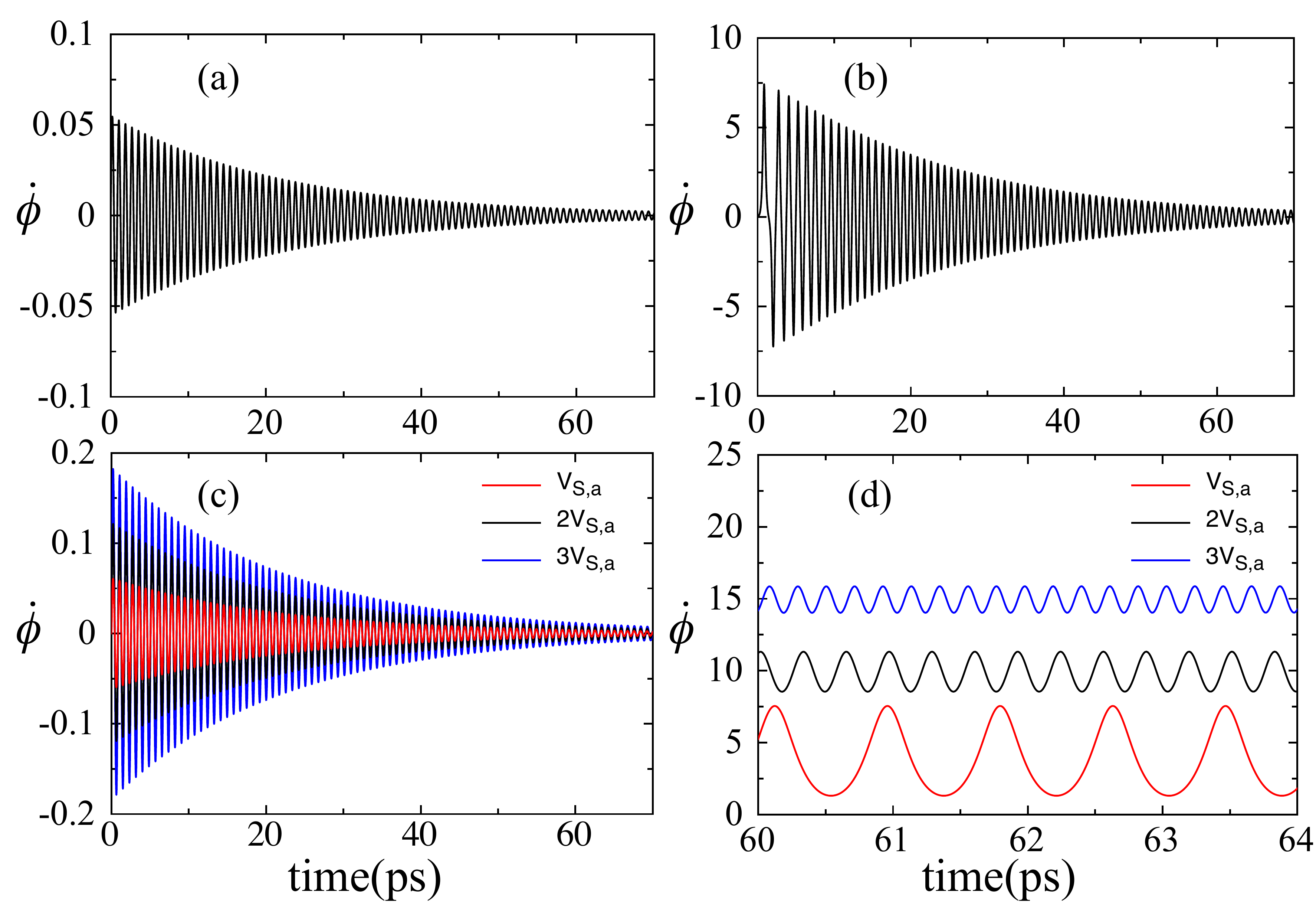} 
\caption{ 
Steady state solution of $\dot{\phi}$ for the 
FM inter-layer exchange ($J < 0$) (a, c), 
and AFM interlayer exchange ($J >0 $) (b, d)
as a function of the spin voltage $V_{S}$. 
We choose a spin injection value $V_{S,a} = 0.031 J$ 
that is greater than the critical spin injection. 
$V_S < V_{S,0}$ for (a) and (b), and $V_{S,a} > V_{S,0}$ for (c) and (d). 
As before, we set $\lambda= 30meV$, $|J|/\lambda = 1/300$ and $\alpha=0.001$.}
\label{fig:figure3}
\end{center} 
\end{figure}

Consider the spin transfer effect in the $\hat{z}$ spin direction at the 
left interface and drop $G_{i}$.
In the presence of a spin accumulation at the left interface, 
Eq.~(\ref{LLG_m}) acquires a spin transfer torque $\tau_{S}$ 
resulting in modified equations for the canonically conjugate fields $(m_{\theta,i},\phi_{i})$. 
Eliminating $m_{\theta}$ from the modified equations, 
the time dynamics of the phase $\phi$ satisfy,
\begin{equation}
\label{phidoubledot}
(1+ \alpha^{2})\ddot{\phi}  + \frac{\hbar \tilde{\alpha} \omega^2}{2J}  \dot{\phi} -  \omega^2 \sin(\phi) = \frac{\omega^2}{2J}  V_{S},
\end{equation} 
where $V_{S} = G^{L}_{r} \mu_{0}/(4\pi)$, 
$\tilde{\alpha} = \alpha + \alpha{'}$ with 
$\alpha{'} = G_{r}/(4\pi)$ is the enhanced damping due to the spin pumping 
at the spacer, and we define a critical spin voltage $V_{S,c} =2J$.  
Here we assume $\alpha{'}$ is small compared to $\alpha$ and take $\alpha \tilde{\alpha} \sim \alpha^2$.
This equation has been extensively studied in the context of superconductivity, and describes the RCSJ model for superconducting Josephson junctions \cite{duzer_principles_1999}.
Based on this similarity, it is prudent to define an effective Stewart-McCumber parameter 
$\beta = 2J(1+\alpha^2)/(\alpha^2(\lambda - J))$, 
which determines over-damped ($\beta \ll 1$) or under-damped ($\beta \gg 1$) junctions.
For typical values of damping in AFMIs $\beta \sim 2J/(\lambda \alpha^2) \gg 1$, 
which corresponds to an under-damped junction where Eq.~(\ref{phidoubledot}) 
must be solved numerically. 

Eq.~(\ref{phidoubledot}) resembles the equation of motion of a particle of mass 
$\hbar^2 (1+\alpha^2)/(2(\lambda - J))$ moving along the $\phi$ axis 
in the presence of an effective tilted washboard potential 
$U(\phi) = 2 J \cos (\phi) - V_{S} \phi$ with a viscous drag force 
$\hbar  \tilde{\alpha}\dot{\phi}$.
The phase dynamics $\dot{\phi}$ in the presence of damping $\alpha = 0.001$
are plotted in Fig.~\ref{fig:figure3}
for various values of a constant spin chemical potential $V_{S}$. 
The steady state solution of $\dot{\phi}$ for both the FM or the AFM 
inter-layer exchange interaction shows the same behavior when 
$V_{S} < V_{S,0}$ (see Fig.~\ref{fig:figure3}(a) \& (b)) 
and different dynamics when $V_{S} > V_{S,0}$ (see Fig.~\ref{fig:figure3}(c) \& (d)), 
where $V_{S,0}=0.031 J \ll V_{S,c}$ depends on $\beta$. 
When the spin chemical potential is small $V_{S} < V_{S,0}$, 
viscous drag dominates the dynamics, and the oscillations in the phase decay 
as a result of the damping for both the FM and AFM inter-layer exchange interaction.
However, if the spin transfer torque induced by $V_{S}$ is sufficiently large, the energy gain due to the spin transfer torque can balance the energy loss due to the damping resulting in a continual rotation of the N\'{e}el vector.
In the language of spintronics, this results from anti-damping like torque due to $V_{S}$ fully compensating the damping torque.
%
%The dynamics of $\dot{\phi}$ at different values of spin chemical potentials $V_{S}$, with $V_{S,0} << V_{S,c} $, are shown %in Fig.3 (c) \& (d) for both the FM and AFM inter-layer exchange interactions.
%
For $J > 0$, which corresponds to the superconducting $\pi$-junction, the system is at an unstable equilibrium point, therefore, a small driving force ($V_{S} \ll 2J$) is enough to induce a full $2 \pi$-rotation of the phase. 
However, for $J < 0$, the system is at an energy minima, so a large driving force $V_{S} \sim 2J $ is required to overcome both the viscous damping force and the force required to push the particle over the hill.

In the over-damped case $\beta \ll 1$, when $V_{S} < V_{S,c}$ a static solution for the phase is allowed $\phi= \sin^{-1}(V/(2J))$ implying $I_{S} =0$.
However, when $V_{S} > V_{S,c}$  only time dependent solutions exist, for $\beta \ll 1$ we can assume $\langle \ddot{\phi} \rangle \sim 0$, solving Eq.~\ref{phidoubledot} gives an oscillation frequency $\omega= 1/ (h \tilde{\alpha}) \sqrt{V^2 - 4J^2}$ for the phase $\phi$, independent of the sign of the inter-layer exchange interaction.
Similar characteristic behavior appears for the case of intermediate damping $\beta \sim 1$, however the critical value of $V_{S,0} = 2\alpha \sqrt{2 |J| \lambda}$ to induce a non-zero steady state $\dot{\phi}$, depends on the damping.
%
%Since Eq.~\ref{phidoubledot} is similar to the RCSJ model of superconducting Josephson junctions, with $I \to I_{S}$ and $V \to V_{S}$ we expect similar non-Ohmic characteristics as those observed in superconducting Josephson junctions.

\begin{figure}[t]
\begin{center}
\includegraphics[width=3.5in]{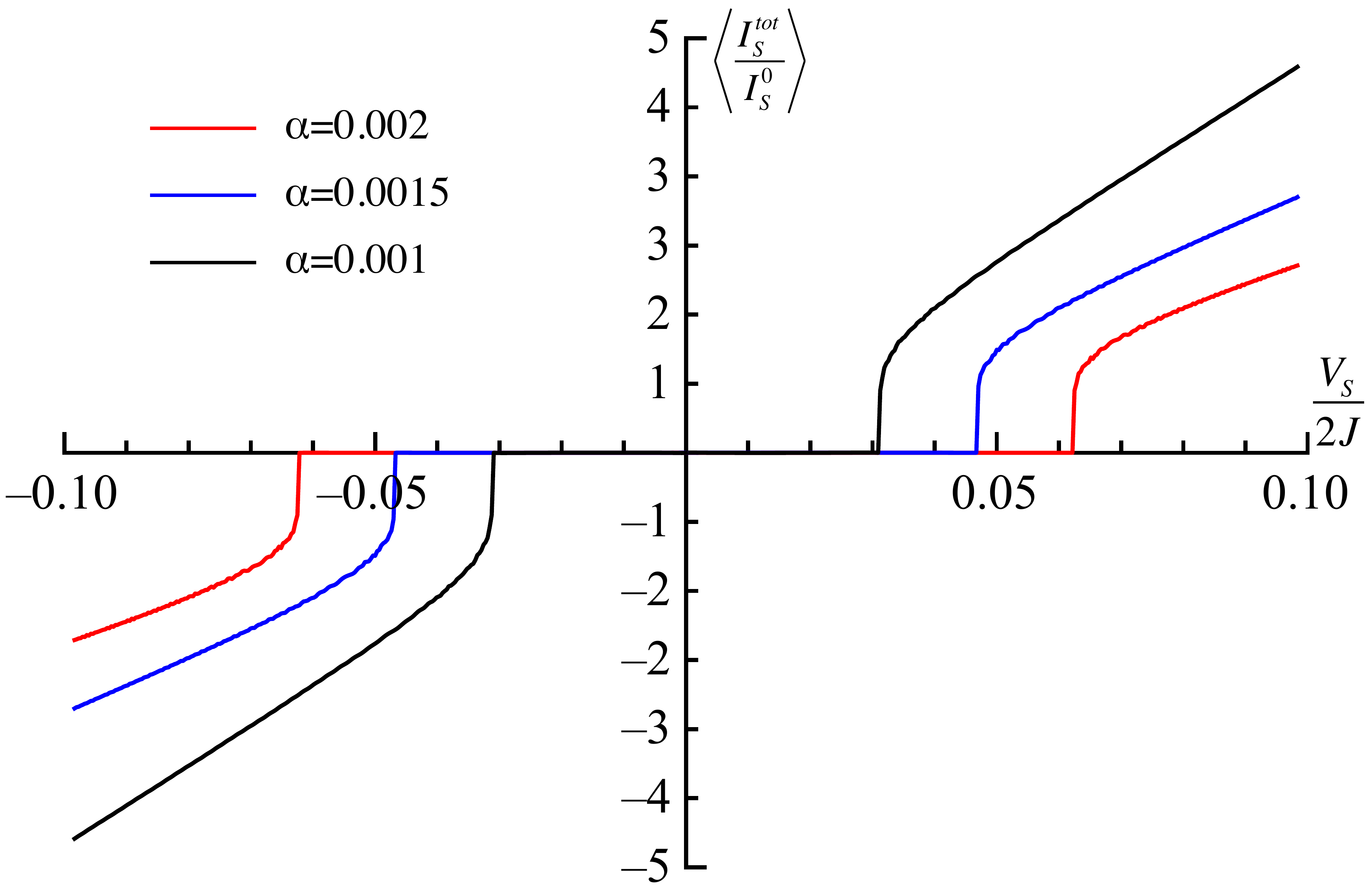} 
\caption{
$I_{S}-V_{S}$ characteristics as a function of the applied spin voltage 
$V_{S}$ for different values of the damping constant with 
$\lambda= 30$ meV and $|J|/\lambda = 1/300$.}
\label{fig:IVC}
\end{center} 
\end{figure}

{\em $I_{S}$-$V_{S}$ characteristics}---The phase dynamics 
associated with both the FM and AFM inter-layer exchange 
interactions result in non-Ohmic $I_{S}$-$V_{S}$ 
characteristics for the AFMI Josephson junctions.
The time averaged value for the spin current $I_{S,av}$ can be determined from 
Eq.~(\ref{spincurrent}) for over-damped and under-damped junctions.
For over-damped AFMI Josephson junctions $\beta \ll 1$, the $I_{S}-V_{S}$ characteristics can be inferred from $I_{S,av} = \hbar G \omega /(4 \pi) $ giving the simple relation,
\begin{equation}
I_{S,av} = \frac{G}{4 \pi \tilde{\alpha}} \sqrt{V_{S}^2- 4 J^2}
\end{equation} 
for $V_{S}> 2J$, which interpolates smoothly between $I_{S,av} =0$ and Ohmic behavior with an effective spin resistance $R_{S} = 4 \pi \tilde{\alpha}/G$. 
The $I_{S}-V_{S}$ characteristics for the under-damped junction 
with an AFM inter-layer exchange ($J >0$) 
are plotted as function of the spin chemical potential $V_{S}$ 
for different values of the damping in Fig.~\ref{fig:IVC}. 
In the under-damped case, the spin current jumps discontinuously from 
$I_{S}=0$ until the spin chemical potential reaches $V_{S,0}$, 
and $V_{S,0}$ is proportional to $ \beta$.
For under-damped junctions with a FM inter-layer exchange ($J < 0$) 
$I_{S,av}=0$ for $V_{S,0} \sim 2J$ where the approximation $|\bf{m}| \ll |\bf{n}|$ breaks down
and requires solutions of LLG equations without any approximations~\cite{inprep}.
The $I_{S}-V_{S}$ characteristics of AFMI spin Josephson junctions 
are different from the $I-V$ characteristics of Josephson junctions in superconductors. 
These differences originate from the spin-current phase relation (see Eq.~\ref{spincurrent}) and 
the role of the spin transfer torque in exchange coupled AFMIs.

{\em Discussion}--- 
This average spin current flowing between the AFMIs can be detected via the
inverse spin Hall effect if the metallic spacer has large spin-orbit 
coupling~\cite{sinova_spin_2015, kajiwara_transmission_2010}. 
There are several ways to induce a spin chemical 
potential~\cite{AllanSSb,takei_superfluid_2014,Brataas_SF}. 
Here we consider spin injection by the spin Hall efect.
To estimate current densities required to drive a spin current, consider two $0.1 \mu \rm{m} \times 0.1\mu \rm{m} \times 0.01\mu \rm{m} $  NiO thin films, with the exchange energy $\lambda = 19.01meV$ and the lattice constant $a=4.17\rm{\AA}$\cite{hutchings_measurement_1972}. 
NiO is an AFM insulating material with an out-of-plane hard axis 
and a much smaller in-plane easy axis (both of them are much smaller than the exchange energy). 
Taking $\alpha=0.007$ and $J = 0.1$ meV we find that a spin chemical potential 
$V_{S,0}=0.039$ meV is required for a spin current $I_{S}= 2.2 \times 10^{-2}$ meV. 
The critical current density can be estimated from the relation 
$V_{S} = \hbar G/(4 \pi e) \theta_{SH} I_c$. 
Taking the spin mixing conductance $g_r$ of NiO of 
$6.9 \times 10^{18} \rm{m^{-2}}$ \cite{cheng_spin_2014}, 
and assuming a $10$nm thick Pt spin current injector with $\theta_{SH} \sim 0.1$, 
we obtain an injection current density $I_c \sim 2.3  \times 10^7 \rm{A}/\rm{cm^2}$.
The induced charge current, $I_{c} = 2 e \theta_{SH} I_{S}/\hbar$ 
through a thin film Pt spacer with $t=1$ nm and  conductivity 
$\sim$ 0.095 $(\mu \Omega {\rm cm})^{-1}$ gives an induced non-local voltage 
$V \sim 0.017 \mu V$ across the Pt spacer.  

Finally, similar oscillations in spin currents can occur across 
exchange coupled easy-plane FMs due to their broken U(1) symmetry. 
Even in the presence of in-plane anisotropy, 
we expect these oscillations to persist as long as 
the spin chemical potential is above the anisotropy energy scale. 
The higher order LLG terms 
do not destroy the spin current oscillations, but they do
affect their detailed dynamics.  
These effects will be discussed elsewhere~\cite{inprep}.
Lastly, the spatial variation in the order parameter, neglected here, can 
nucleate spin solitons or instantons within the junction, 
which can lead to a Fraunhofer-like interference patterns in 
the non-local voltage similar to the behavior of critical 
super-currents in superconducting Josephson junctions. 
\\
\\

\noindent
{\em Acknowledgement:}
This work was supported as part of the Spins and Heat in
Nanoscale Electronic Systems (SHINES)
an Energy Frontier Research Center funded by the U.S. Department of Energy,
Office of Science, Basic Energy Sciences under Award \#DE-SC0012670.
Initial analytical work was also supported by the NSF ECCS-1408168.

%\bibliography{Spin_Josephson_draft}


\end{document}